\documentclass{svproc}

\usepackage{url}
\usepackage{graphicx}
\usepackage{amsmath}
\usepackage{amssymb}
\usepackage{epstopdf}

\newcommand{\be}{\begin{equation}}
\newcommand{\ee}{\end{equation}}
\newcommand{\rvec}{\ensuremath{\boldsymbol{r}}}
\newcommand{\Rvec}{\ensuremath{\boldsymbol{R}}}

\begin{document}
\mainmatter              
\title{Efimov physics beyond three particles}
\titlerunning{Efimov physics beyond three particles}
\author{Betzalel~Bazak\inst{1}}
\authorrunning{Betzalel Bazak} 
\institute{The Racah Institute of Physics, The Hebrew University, 9190401,
  Jerusalem, Israel\\
\email{betzalel.bazak\,@\,mail.huji.ac.il}}
\maketitle              

\begin{abstract}
  Efimov physics originally refers to a system of three particles. 
  Here we review recent theoretical progress seeking for manifestations of
  Efimov physics in systems composed of more than three particles.
  Clusters of more than three bosons are tied to each Efimov trimer,
  but no independent Efimov physics exists there beyond three bosons.
  The case of a few heavy fermions interacting with a lighter atom is also
  considered, where the mass ratio of the constituent particles plays a
  significant role.
  Following Efimov's study of the $(2+1)$ system, the $(3+1)$ system was
  shown to have its own critical mass ratio to become Efimovian.
  We show that the $(4+1)$ system becomes Efimovian at a mass ratio which
  is smaller than its sub-systems thresholds, giving a pure five-body
  Efimov effect.
  The $(5+1)$ and $(6+1)$ systems are also discussed, and we show the absence of $6-$ and $7-$body Efimov physics there. 
  \keywords{Efimov physics, ultracold atoms, universal physics}
\end{abstract}

\section{Introduction}

Universal aspects of few-body systems with large scattering length have
attracted attention in recent years from both theory and experiment
perspectives \cite{BraHam06}. 
Universality occurs when there is a large separation between the scale of the
underlying physics and the scale of the phenomena observed.
For example, if the inter-particle interaction range is much shorter than the
spatial extent of the wave function governed by the scattering length $a$, most
of the time the particles will be out of the potential range and therefore
not sensitive to its fine details. 

A few examples of relevant systems come to mind. 
In low-energy nuclear physics the scattering length of the singlet and triplet
channels are $a_s\approx -23.4$ fm and $a_t\approx 5.42$ fm,
while the long-range part of the nucleonic interaction, determined by the
pion mass, is
$R \approx \hbar / m_\pi c \approx 1.4$ fm.

Larger scale separation can be found in $^4$He atoms. Here the He-He
scattering length is $a\approx 90$ \AA, while the Van Der Waals interaction
range is  $r_{vdW}\approx 5.4$ \AA.

Another interesting case is ultracold atoms near Feshbach resonance.
Here the scattering length can be tuned to arbitrary value using, for example,
external magnetic field, 
$$ a(B) = a_{bg}\left(1+\frac{\Delta}{B-B_0}\right)\,. $$

A fascinating effect was predicted by Efimov \cite{Efimov70} for three
identical bosons with resonating interaction:
the existence of infinite tower of bound trimers.
For a recent review see \cite{NaiEnd17}.

In this paper, we study Efimov physics beyond three particles. We start with
a short review of universal features and Efimov physics in three identical
bosons and in the $(2+1)$ system, which is a mixture of two identical fermions
and distinguishable particle.
Then we go beyond three particles and discuss the $N>3$ identical bosons
system, as well as the $(N+1)$ systems with $N \le 6$. 
 
\section{Methods and results}

In order to study universality, one would like to neglect the system-specific
details and concentrate on the universal features. 

To do so one could use the zero-range limit, i.e. eliminate the
spatial extent of the potential while applying the Bethe-Peierls boundary
condition when two particles touch each other,
\be\label{BC}
\frac{\partial \log (r_{ij} \psi)}{\partial r_{ij}}
\xrightarrow[r_{ij} \to 0]{}-\frac{1}{a}
\ee
where $r_{ij}=|\rvec_j-\rvec_i|$ is the distance between any pair of interacting
particles.

\subsection{The universal dimer}

A trivial example for universality is the existence of universal dimer
composed of two identical bosons of mass $m$ for $a>0$.
Working in the center-of-mass frame and taking the zero-range limit,
one has to solve the free Schr\"odinger equation for the relative coordinate
$r$ and to apply the Bethe-Peierls boundary condition (Eq. \ref{BC}) at zero,
giving for the bound state $\psi(r) \propto exp(-r/a)/r$ corresponds to
an energy of $-1/ma^2$, where here and thereafter $\hbar$ is set to $1$. 

This prediction is indeed valid for the three examples mentioned above. 
The deuteron binding energy, 2.22 MeV, is fairly close to the universal
prediction $1/ma_t^2\approx 1.4$ MeV.
The $^4$He atoms dimer binding energy was measured recently to be about
$1.76(15)$ mK \cite{Zel16}, where the universal prediction is
$1/ma^2\approx 1.48$ mK.
Since the next correction is of order of $r_0/a$, where $r_0$ is the effective
range, one would expect the universal prediction to be even better with
ultracold atoms, and indeed this is the case \cite{ThoHodWie05}. 

\subsection{Efimov physics in three identical bosons} 

Adding another identical boson, the situation is changed dramatically,
as Efimov has shown \cite{Efimov70}.

To see that one can start from the Faddeev equation for zero-range potential,
and then transform to hyperspherical coordinates.
In the unitary limit $a\longrightarrow\infty$,
the energy is then determined by one-dimensional equation for the hyper-radius,
$\rho^2 \propto r_{12}^2+r_{23}^2+r_{13}^2$, 
\be\label{hyperRadial}
  \left(-\frac{d^2}{d\rho^2}+\frac{s^2-1/4}{\rho^2}\right)R(\rho)
  =ER(\rho),
\ee
where $s$ is the eigenvalue of the corresponding
hyper-angular equation. Eq. (\ref{hyperRadial}) has two interesting features.
First, the effective three-body potential
has long range $\propto \rho^{-2}$, in contrast to the zero-range two-body
interaction we start with.
Second, it exhibits scale invariance, therefore if $R(\rho)$
is a solution with the corresponding energy $E$, $R(\lambda \rho)$ is also
solution with the energy $\lambda^2 E$ for arbitrary $\lambda$.

At small $\rho$, $E$ can be neglected, and the solution for
Eq. (\ref {hyperRadial}) is $R_\pm(\rho) \propto \rho^{1/2 \pm s}$.
The solution behavior is therefore determined by $s$.
For $s^2>0$ the solution can be set to $R_+(\rho)$, 
while for $s^2<0$ the solution is a combination of two oscillating
functions, whose relative phase is still needed to be fixed.

In the latter case the effective potential is attractive and  
one faces fall of a particle to the center of $\rho^{-2}$ potential,
i.e. the energy here is not bound from below \cite{Thomas35}.

Introducing three-body potential barrier at some finite $\rho_0$
saves us from this collapse by setting the system ground state.
The scale invariance is now broken into discrete scale invariance,
with $\lambda_n=e^{-\pi n/|s|}$, and therefore the energies are quantized,
giving infinite series of bound states with geometric-series spectrum
$E_n=E_0 e^{-2\pi n/|s|}$.
Here $\rho_0$ is a three-body parameter, which sets the ground state energy
and also fixes the relative phase of $R_\pm$. $s$ is the scale factor
which governs the scaling characters of the energies and the wave functions.
For three identical boson it has the value $s=1.00624\,i$.

This prediction had to wait about four decades before its verification
in ultracold gases experiments,
where particle loss from the trap is a three-body process,
showing a significant signal when new Efimov state is formed. 
Studies of loss features of ultracold $^{133}$Cs \cite{Kra06}, 
$^{39}$K \cite{zac09}, and $^{7}$Li \cite{gro09,pol09} gases gave
the first experimental verification for Efimov physics.

The existence of Efimov trimers in $^4$He atoms was predicted long ago
\cite{LimDufDam77}, where due to the finite scattering length
only two trimers should exist.
Only recently the excited trimer was seen experimentally \cite{Kun15},
giving another verification for Efimov's prediction.

\subsection{Efimov physics beyond three identical bosons}

Shortly after Efimov original paper, Amado and Greenwood have claimed that 
there is no Efimov effect for four or more particles \cite{AmaGre73}.

However, three-body Efimov physics has a footprint in the four body system, 
were two tetramers are tied to each Efimov trimer \cite{HamPla07}, 
a prediction which was verified in ultracold atoms experiments
\cite{SteDinGre09}. 
The tetramer and trimer energies are correlated, 
similar to the correlation between triton and alpha binding energies
known as the Tjon line \cite{tjon75,PlaHamMei05}.

Larger clusters of identical bosons also exist,
and their energies are correlated to the trimer energy,
therefore not showing independent $N$-boson Efimov physics
\cite{BazEliKol16}. See, however, \cite{Baz19}. 
   
\subsection {Mass imbalanced fermionic mixtures}

Another system relevant to Efimov physics is a mixture of identical fermions
and distinguishable particle with different mass.

Consider two heavy atoms with mass $M$ interacting with a light atom with
mass $m$.
First, we apply the Born-Oppenheimer approximation, valid in the limit
of $M \gg m$ \cite{FonRedSha79,Petrov12}.
Here the motion of the light particle is first solved assuming the heavy
particles position is fixed at $\pm R/2$, giving 
\be
 \psi_{\Rvec}^\pm(\rvec) \propto \frac{e^{-\kappa(R)|\rvec-\Rvec/2|}}{|\rvec-\Rvec/2|}
 \pm \frac{e^{-\kappa(R)|\rvec+\Rvec/2|}}{|\rvec+\Rvec/2|}
\ee
where $\rvec$ is the light particle position. Applying the boundary condition
(Eq. \ref{BC}) gives
\be
\kappa_\pm(R)\mp\frac{e^{-\kappa_\pm(R)R}}{R}=1/a.
\ee
The energy of the light atom $\epsilon_\pm(R)=-\kappa_\pm^2(R)/2m$ is
then considered as an effective potential between the heavy atoms.
The minus state corresponds to repulsive effective potential,
while the plus state induces attractive potential,
\be
\epsilon_+(R)\approx
  \begin{cases}
    -\frac{0.16}{mR^2} & R/a \ll 1 \\
    -\frac{1}{2m}\left(\frac{1}{a^2}+\frac{\exp(-R/a)}{aR}\right) & R/a \gg 1
  \end{cases}
\ee

The heavy-particles equation for $R\ll a$ is therefore identical to Eq. (\ref{hyperRadial}),
replacing $\rho$ by $R$. Here $s^2=l(l+1)-0.16M/m+1/4$ for angular momentum $l$.

In the bosonic case, the ground state has $l=0$,
giving purely attractive $-1/R^2$ effective potential, and therefore
Efimov physics.

\begin{figure}\begin{center}
  \includegraphics[]{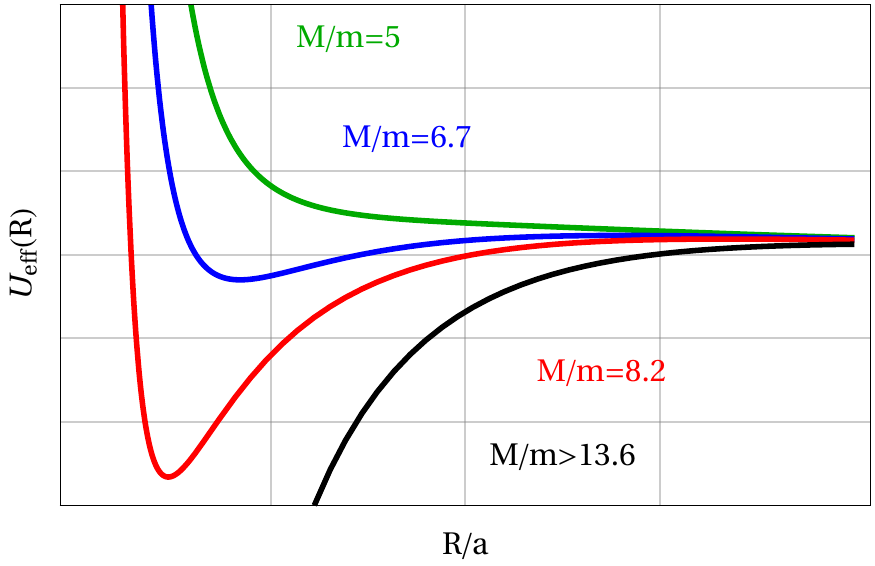} \end{center}
  \caption{The effective potential between heavy particles induced by the light
    particle as a function of their distance, for different mass ratios in the
    Born-Oppenheimer approximation.
    For a small mass ratio, the effective interaction is repulsive (green).
    As the mass ratio increases $p$-wave resonance occurs (blue),
    then the potential well becomes deep enough to support bound trimer (red),
    and finally, the attraction wins and Efimov states emerge (black).
    \label{BO}}
\end{figure}

In the fermionic case, the permutation symmetry dictates odd angular
momentum and the ground state has $l=1$.
The centrifugal barrier $l(l+1)/MR^2$ therefore competes with
the $-1/mR^2$ attraction, where the competition is governed by the mass ratio.
Fig. \ref{BO} shows the effective potential for various mass ratios $M/m$. 

This simple picture indeed catches the physics here.
For a small mass ratio, the effective potential is repulsive, and no bound
trimer exists. 
As the mass ratio increases, the potential becomes more attractive,
and a $p$-wave resonance occurs. 
Indeed this resonance was found in ultracold $^{40}$K-$^6$Li mixture
\cite{Jag14}.
For larger mass ratio the potential well is deep enough to support a
universal $1^-$ bound state \cite{KarMal07}.
For even larger mass ratio the system becomes Efimovian \cite{Efimov73}.

To proceed beyond this approximation, it is convenient to follow Skorniakov and
Ter-Martirosian formalism \cite{STM57,Pri11}.
Here instead of solving the Schr\"odinger equation, one utilizes the zero-range
potential to get an integral equation.

For the $(N+1)$ case, the STM equation in momentum space is
\cite{BazPet17},
\begin{equation}\label{STM}
\frac{1}{4\pi}\left(\frac{1}{a}-\kappa \right)F(\mathbf q_1,\ldots, \mathbf q_{N-1})=\int \frac{d^3 q_N}{(2\pi)^3} 
\frac{\sum_{i=1}^{N-1}F(\mathbf q_1,\ldots,\mathbf q_{i-1},\mathbf q_N,\mathbf q_{i+1},\ldots,\mathbf q_{N-1})}
{-2\mu E +\frac{\mu}{M} \sum_{i=1}^N q_i^2 + \frac{\mu}{m} \left( \sum_{i=1}^{N} \mathbf q_i \right)^2 },
\end{equation}
where  $\mu=Mm/(M+m)$ is the reduced mass.
The function $F(\mathbf q_1,\ldots, \mathbf q_{N-1})$ can be considered as
the relative wave function of $N-1$ heavy atoms with momenta
${\mathbf q}_1,\ldots,{\mathbf q}_{N-1}$ and a heavy-light
pair, the momentum of which equals $-\sum_{i=1}^{N-1} {\mathbf q}_i$. Here 
\be
\kappa=\sqrt{-2\mu E+\frac{\mu}{M} \sum_{i=1}^{N-1} q_i^2 +
  \frac{\mu}{M+m} \left( \sum_{i=1}^{N-1} \mathbf q_i \right)^2}.
\ee

\begin{figure} \begin{center}
  \includegraphics[]{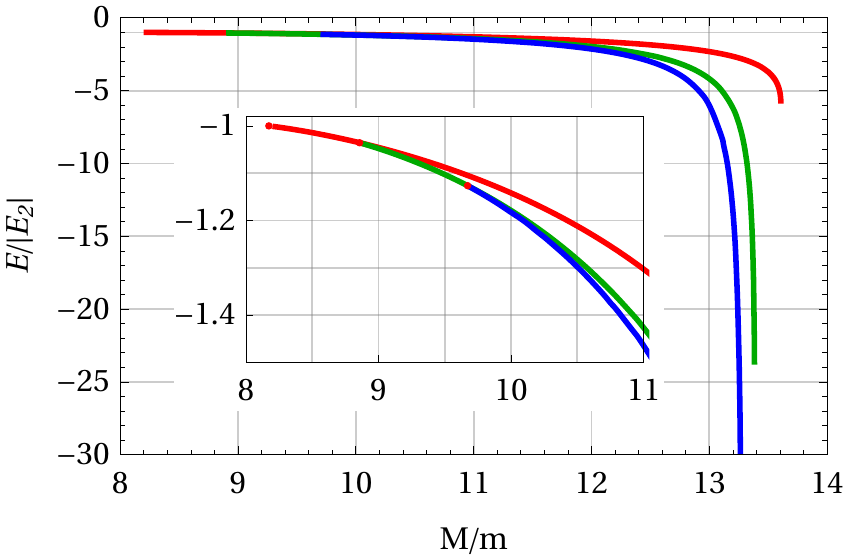} \end{center}
  \caption{The energies of the universal $(N+1)$ states in units of the dimer
    binding energy, as a function of the mass ratio.
    Shown are results for the $1^-$ $(2+1)$ state (red),
    the $1^+$ $(3+1)$ state (green), and the $0^-$ $(4+1)$ state (blue).
    The inset shows zoom-in on the thresholds region.
    Adapted from \cite{BazPet17}
    \label{energies}}
\end{figure}

For a specific value of total angular momentum and parity $L^\pi$, the
equation can be further simplified.
For the $(2+1)$ case the relevant symmetry for both universal trimer and
Efimov state is $1^-$.
The function $F$ therefore takes the form
$ F(\mathbf q)=f(q) \hat{ \mathbf z} \cdot \hat{ \mathbf q} $,
leaving one-dimensional equation which can be easily solved for $E$,
as can be seen in Fig. \ref{energies}.
The universal trimer binding threshold could thus be obtained,
$M/m=8.173$, in agreement with the results obtained in Ref. \cite{KarMal07} in
the hyperspherical formalism.

To find the mass ratio where the system becomes Efimovian one would like
to calculate the scale factor $s$ which approaches zero at that point.
For that, one can calculate the large-$q$ asymptote of $f$, which has the form 
\be
f(q)\propto q^{-2-s}.
\ee
Solving Eq. (\ref{STM}) for $f$ and fitting the results to extract $s$,
the Efimov threshold can be found at $M/m=13.607$, 
in agreement with the result of Ref. \cite{Efimov73}.

An interesting alternative, which may be more suitable for larger systems,
is to utilize the mapping between the free-space system with finite $a$ and
the trapped system at unitarity
\cite {Tan04,WerCas06},
whose energy is 
\be \label{HO}
E=\hbar \omega (s+2n+1)
\ee
where $\omega$ is the trap frequency, $s$ is the same scale factor and $n$
counts hyper-radial excitations. 
Hence, one can extract the scale factor $s$ from the trapped energies.

Now that we have built our toolbox, we can face an interesting question:
how many heavy fermions can be bound by a single light atom?

For the $(3+1)$ case the relevant symmetry is $1^+$, and $F$ takes the form
$
F(\mathbf q_1,\mathbf q_2)=f(q_1,q_2,\hat{\mathbf q}_1\cdot \hat{\mathbf q}_2)
\hat{ \mathbf z} \cdot \hat {\mathbf q}_1 \times \hat{\mathbf q}_2, 
$
leaving a three-dimensional integral equation which can be solved in
deterministic method. 
It was shown that a universal $1^+$ tetramer exists for a mass ratio
$M/m \gtrsim 9.5$ \cite{Blu12}.
Moreover, an $1^+$ Efimov states exist above $M/m>13.384$ \cite{CasMorPri10}.

Adding another fermion, the relevant symmetry is $0^-$, therefore
$
F(\mathbf q_1,\mathbf q_2,\mathbf q_3)=f(q_1,q_2,q_3,\hat{\mathbf q}_1\cdot \hat{\mathbf q}_2,\hat{\mathbf q}_1\cdot \hat{\mathbf q}_3,\hat{\mathbf q}_2\cdot \hat{\mathbf q}_3) \hat{\mathbf q}_1 \cdot \hat{\mathbf q}_2 \times \hat {\mathbf q}_3, 
$
but the resulting six-dimensional integral equation is too hard to be solved
with conventional method.
Hence a novel method, which we call the STM-DMC method, is introduced \cite{BazPet17,BazPet18}, where $f$ is treated as density probability
function for so-called walkers, whose stochastic dynamics is governed in
such a way that their detailed-balance condition is Eq. \ref{STM}.
Given $E$, $a$ is than changed in each iteration to keep the walkers' number
constant. 

Using this method, Eq. (\ref{STM}) can be solved for the $(2+1)$, $(3+1)$,
and $(4+1)$ cases, getting both energies and scale factors. 
Fig. \ref{energies} shows the energies of the universal states in these
systems, its inset focuses on their thresholds. 
Known results are reproduced, i.e. the thresholds for the
$(2+1)$ universal trimer and Efimov states.
Moreover, we can locate better the threshold of the $(3+1)$ universal tetramer
to be $8.862$ \cite{BazPet17},
and confirm the threshold for four-body Efimov states \cite{CasMorPri10}.

\begin{figure} \begin{center}
  \includegraphics[]{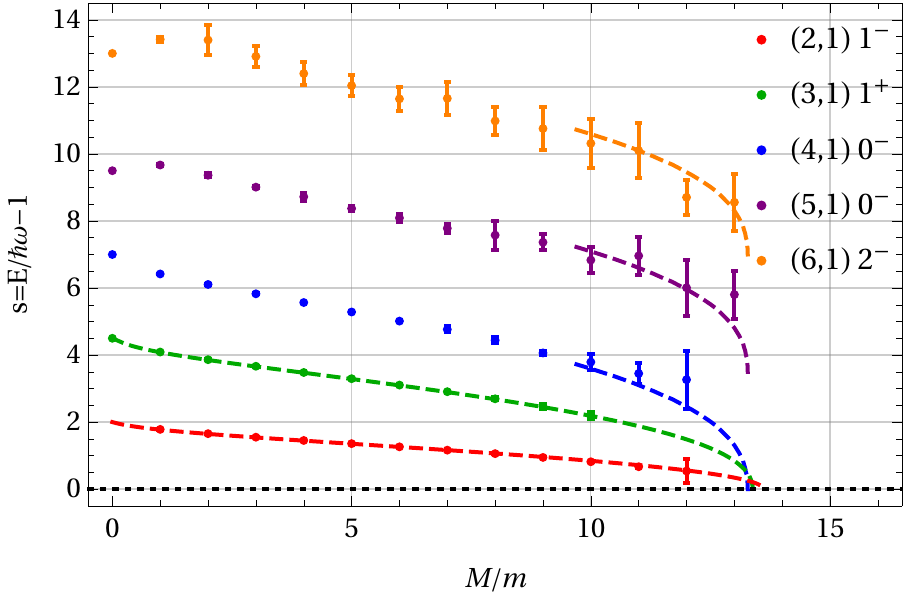} \end{center}
  \caption{The scale factor as extracted from the energies of the $(N+1)$
    state in a harmonic trap at unitarity, as a function of the mass ratio.
    The Efimov limit corresponds here to $s=0$. 
    Circles stand for the results extrapolated from finite-range gaussian
    potential to the zero-range limit.
    Dashed curves are for results acquired directly in the zero-range limit,
    by solving the STM equation on a grid (for $(2+1)$ and $(3+1)$ systems)
    or with stochastic method (for the $(4+1)$ case).
    The Dashed curves for the $(5+1)$ and $(6+1)$ cases taken from
    the $(4+1)$ case with appropriate shift, showing no Efimov
    effect exists in these cases.
    \label{s}}
\end{figure}

We can now explore the \emph{terra incognita} $(4+1)$ system.
Here we find a $0^-$ universal pentamer as well as $0^-$ Efimov states. 
In Fig. \ref{energies} we also plot the energies for the 
universal $0^-$ pentamer, showing it is bound for mass ratio above $9.672$.
In Fig. \ref{s} we show the scale factor for this system, showing
Efimov $0^-$ states emerge here above $M/m=13.279$ \cite{BazPet17}.

The different threshold for the $(N+1)$ states are summarized in Tab. \ref {thresholds}.

\begin{table}
  \caption{The thresholds for universal and Efimov states in the $(N+1)$ systems.}
  \label{thresholds}
  \begin{center}
\vspace{0.3cm}
{\renewcommand{\arraystretch}{1.25}%
\begin{tabular}
{c@{\hspace{4mm}} c@{\hspace{3mm}} c@{\hspace{4mm}} c}
\hline \hline
system & $L^\pi$ & universal state & Efimov state \\
\hline
2+1 & $1^-$ & 8.173 & 13.607 \\
3+1 & $1^+$ & 8.862 & 13.384 \\
4+1 & $0^-$ & 9.672 & 13.279 \\
\hline
5+1 & $0^-$ & ?     & ---    \\
6+1 & $2^-$ & ?     & ---    \\
\hline \hline
\end{tabular}}
\end{center}
\end{table}

We see that the $(N+1)$ systems with $N=2,3$ and $4$ exhibit similar behavior,
showing pure $N+1$-body Efimov physics. Does this pattern continue
for $N \ge 5$? 

The relevant symmetry for the $(5+1)$ ground state is $0^-$, signaling that the
additional fermion populates an excited $s$-shell, which has a radial node.
This causes the stochastic method to suffer from a sign problem.
Hence we choose here another approach, which is to extract the scale factor from
the energies in a harmonic trap. 

These energies were calculated using a Gaussian potential with a finite
range $R_0$,
\be
V(r)=-V_0 \,e^{-\frac{r^2}{2R_0^2}}
\ee
where the results are extrapolated to the zero-range limit
$R_0\longrightarrow 0$.
The $(N+1)$-body Schr\"odinger equation is solved using a basis of correlated
gaussians, chosen by the stochastic variational method \cite{SVM}.
Using Eq. (\ref{HO}), the scale factor is then extracted \cite{Trap}.

In Fig. \ref{s} we show the $(N+1)$ system scale factor for $N \le 6$.
Efimov threshold here is signaled by $s=0$.
Results obtained with other methods are also shown.
Indeed the scale factors for $N=2,3$ and $4$
hit zero at the Efimovian threshold.
However, no sign for Efimov physics is found in the
$N=5$ and $6$ systems. 

\section{Conclusion}
Efimov physics beyond three particles is studied here. For identical bosons, no independent Efimov effect exists beyond three particles, although bosonic
clusters are tied to each Efimov trimer.
For the $(N+1)$ case of $N$ identical fermions interact with distinguishable
particle, Efimov states occur for mass ratio exceeds the relevant threshold
for the $(2+1)$, $(3+1)$, and $(4+1)$ systems. 
However, no Efimov state exists for the $(5+1)$ and $(6+1)$ systems. 

\section*{Acknowledgements}
I would like to thank Dmitry Petrov for useful discussions and
communications.


\begin{thebibliography}{99}
%
\bibitem{BraHam06} Braaten E. and Hammer H.-W.:
  Universality in few-body systems with large scattering length,
  Phys. Rep. {\bf 428}, 258 (2006).

\bibitem{Efimov70} Efimov V.:
  Energy levels arising from resonant two-body forces in a three-body system,
  Phys. Lett. B {\bf 33}, 563 (1970)

\bibitem{NaiEnd17} Naidon P. and Endo S.:
  Efimov physics: a review, 
  Rep. Prog. Phys. {\bf 80}, 056001 (2017).

\bibitem{Zel16}  Zeller S. \textit{et al.},
  Imaging the He$_2$ quantum halo state using a free electron laser,
  Proc. Natl. Acad. Sci. U.S.A. {\bf 113}, 14651 (2016).

\bibitem{ThoHodWie05} Thompson S. T., Hodby E., and Wieman C. E.:
  Ultracold Molecule Production via a Resonant Oscillating Magnetic Field,
  Phys. Rev. Lett. {\bf 95}, 190404 (2005).

\bibitem{Thomas35} Thomas H.:
   The interaction between a neutron and a proton and the structure of h$^3$,
   Phys. Rev. {\bf 47}, 903 (1935)

 \bibitem {Kra06} Kraemer T., Mark M., Waldburger P., Danzl J. G., Chin C.,
   Engeser B., Lange A. D., Pilch K., Jaakkola A., N\"agerl H.-C. and Grimm R.:
   Evidence for Efimov quantum states in an ultracold gas of caesium atoms,
   Nature {\bf 440}, 315 (2006).   
   
 \bibitem {zac09} Zaccanti M., Deissler B., D'Errico C., Fattori M.,
   Jona-Lasinio M., Müller S., Roati G., Inguscio M., Modugno G.:
  Observation of an Efimov spectrum in an atomic system, 
  Nature Phys. {\bf 5}, 586 (2009).

\bibitem {gro09} Gross N., Shotan Z., Kokkelmans S., and Khaykovich L.:
  Observation of universality in ultracold $^7$Li three-body recombination,
  Phys. Rev. Lett. {\bf 103}, 163202 (2009).

\bibitem {pol09} Pollack S. E., Dries D., and Hulet R. G.:
  Universality in three- and four-body bound states of ultracold atoms,
  Science  {\bf 326}, 1683 (2009).

\bibitem{LimDufDam77} Lim T. K., Duffy S. K. and Damert W. C.:
  Efimov state in the $^4$He trimer,
  Phys. Rev. Lett. {\bf 38}, 341 (1977).

\bibitem{Kun15}
  Kunitski M. \textit{et al.}:
  Observation of the Efimov state of the helium trimer,
  Science {\bf 348}, 551 (2015).

\bibitem{AmaGre73} Amado R. D. and Greenwood F. C.:
  There is no Efimov effect for four or more particles,
  Phys. Rev. D {\bf 7}, 2517 (1973).

\bibitem{HamPla07} Hammer H.-W. and Platter L.:
  Universal properties of the four-body system with large scattering length,
  Eur. Phys. J. A {\bf 32}, 113 (2007).

\bibitem{SteDinGre09}
  von Stecher J, D'Incao J P, and Greene C H: 
  Signatures of universal four-body phenomena and their relation to the Efimov
  effect,
  Nat. Phys. {\bf 5}, 417 (2009).

\bibitem{tjon75} Tjon J. A.:
  Bound states of $^4$He with local interactions,
  Phys. Lett. B {\bf 56}, 217 (1975).

\bibitem{PlaHamMei05} Platter L., Hammer H.-W., Meissner U.-G.:
  On the correlation between the binding energies of the triton and the
  alpha-particle,
  Phys. Lett. B {\bf 607}, 254 (2005).

\bibitem{BazEliKol16} Bazak B., Eliyahu M., and van Kolck U.:
  Effective Field Theory for Few-Boson Systems,
  Phys. Rev. A, {\bf 94}, 052502 (2016).

\bibitem{Baz19} Bazak B., Kirscher J., K\"onig S., Pav\'on Valderrama M.,
  Barnea N., and van Kolck U.:
  The four-body scale in universal few-boson systems,
  arXiv:1812.00387.
  
\bibitem{FonRedSha79}  Fonseca A. C., Redish E. F., and Shanley P. E.:
  Efimov effect in an analytically solvable model,
  Nucl. Phys. A {\bf 320}, 273 (1979).

\bibitem{Petrov12} Petrov D. S.:
  The few-atom problem,
  in Many-body physics with ultra-cold gases: Lecture Notes of the 2010
  Les Houches Summer School, {\bf 94}, (2012).

\bibitem{Jag14} Jag M., Zaccanti M., Cetina M., Lous R. S., Schreck F.,
  Grimm R., Petrov D. S. and Levinsen J.:
  Observation of a strong atom-dimer attraction in a mass-imbalanced
  Fermi-Fermi mixture,
  Phys. Rev. Lett. {\bf 112}, 075302 (2014).

\bibitem{KarMal07} Kartavtsev O. I. and Malykh A. V.:
  Low-energy three-body dynamics in binary quantum gases,
  J. Phys. B: At. Mol. Opt. Phys. {\bf 40}, 1429 (2007).

\bibitem{Efimov73} Efimov V.:
  Energy levels of three resonantly interacting particles,
  Nucl. Phys. A {\bf 210}, 157 (1973).

\bibitem{STM57} Skorniakov, G. V. and Ter-Martirosian, K. A.:
  Three body problem for short range forces. I. Scattering of low energy
  neutrons by deutrons,
  Sov. Phys. JETP {\bf 4}, 648 (1957).

\bibitem{Pri11} Pricoupenko L.:
  Isotropic contact forces in arbitrary representation: Heterogeneous few-body
  problems and low dimensions,
  Phys. Rev. A {\bf 83}, 062711 (2011).

\bibitem{BazPet17} Bazak B. and Petrov D. S.:
   Five-Body Efimov Effect and Universal Pentamer in Fermionic Mixtures,
   Phys. Rev. Lett. {\bf 118}, 083002 (2017).

\bibitem{Tan04} Tan S.:
  Short range scaling laws of quantum gases with contact interactions,
  arXiv:cond-mat/0412764 (2004).

\bibitem{WerCas06} Werner F. and Castin Y.:
  Unitary gas in an isotropic harmonic trap: Symmetry properties and
  applications,
  Phys. Rev. A {\bf 74}, 053604 (2006).

\bibitem{Blu12} Blume D.:
  Universal four-body states in heavy-light mixtures with a positive
  scattering length,
  Phys. Rev. Lett. {\bf 109}, 230404 (2012).
  
\bibitem{CasMorPri10} Castin Y., Mora C., and Pricoupenko L.:
  Four-body Efimov effect for three fermions and a lighter particle,
  Phys. Rev. Lett. {\bf 105}, 223201 (2010).

\bibitem{BazPet18} Bazak B. and Petrov D. S.:
   Energy of N two-dimensional bosons with zero-range interactions,
   New J. Phys. {\bf 20}, 023045 (2018). 

\bibitem{SVM} Suzuki Y. and Varga K.:
  Stochastic variational approach to quantum-mechanical few-body problems,
  Springer 1998.
       
\bibitem{Trap} Bazak B.:
   Mass-imbalanced fermionic mixture in a harmonic trap,
   Phys. Rev. A {\bf 96}, 022708 (2017).

\end{thebibliography}
\end{document}